\documentclass{article}
\usepackage[accepted]{vietnam} 
\usepackage{natbib}
\usepackage{graphicx}  
      \def\new#1 {{\bf #1 }}
      \def\cut#1 {\sout{#1} }

\def\degr {\hbox{$^\circ$}}

\def\Msol {$\hbox{M}_\odot$}

\def\simgreat{\mathbin{\lower 3pt\hbox
     {$\rlap{\raise 5pt\hbox{$\char'076$}}\mathchar"7218$}}}
\def\simless{\mathbin{\lower 3pt\hbox
     {$\rlap{\raise 5pt\hbox{$\char'074$}}\mathchar"7218$}}}

\begin{document}
\twocolumn[
\title{Magnetic Fields in Star--Forming Filaments in Different Environments}
\titlerunning{Magnetic Fields in Filaments}
\author{Thushara Pillai}{tpillai.astro@gmail.com}
\address{Max--Planck--Institut f\"ur Radioastronomie (MPIfR),  Bonn, Germany}

\vskip 0.5cm 
]

\begin{abstract}
Cold, dense filaments, some appearing as infrared dark clouds, are the nurseries of  stars. Tremendous progress in terms of temperature, density distribution and gas kinematics has been made in understanding the nature of these filaments. However,  very little is known about the role played by magnetic fields in the evolution of these filaments. Here, I summarize the recent observational efforts and  ongoing projects (POLSTAR survey) in this direction. 
\end{abstract}

\section{Introduction}
Understanding the dependence of star--formation on galactic
environment is one of core objectives of this conference. One of the most striking features of the ISM as revealed by
the dust emission surveys with Herschel is the ubiquity of filaments
at all spatial scales \citep{andre2014a}. The densest and coldest of
such filaments now known as Infrared Dark Clouds (IRDCs) appear in
silhouette against the Galactic mid-infrared (MIR) backgrounds. Their
mid-infrared properties suggested that IRDCs might be cold, with high
masses (high mid-IR opacity).  Since their discovery, cores within
IRDCs have also been proposed to be the missing link 
in the evolutionary paradigm for high-mass stars --  massive pre-stellar cores
(\citealt{carey1998:irdc,rathborne2006:irdc_bolometer,ragan2009:irdcs}.)

A systematic survey of the Spitzer data has revealed several $10^3$ of
such IRDCs strewn along the galactic plane \citep{peretto2009}. Most of the mass contained by IRDCs is
concentrated in just a few $10^2$ clouds, but that these clouds might
form the main mass reservoir for future star formation
\citep{kauffmann2010c}. The most massive IRDCs have masses exceeding
$10^4$\,\Msol \citep{kainulainen2011:irdc,butler2009}. A critical
question in ISM research is how such large masses are being swept up
into forming IRDCs?  Clearly, the answer lies in understanding the two
dominant pressures in the cold dense phase: Magnetic field and
turbulence. Turbulence can be characterized by wide-field mapping of
IRDCs in tracers sensitive to different densities and
temperatures. Several large scale surveys with single dish telescopes
and smaller scale surveys with interferometers have been undertaken in
the recent years to achieve this goal
\citep{foster2011,ragan2011:vla,tan2014}. Thanks to such surveys we
know that turbulence is supersonic. We also know that the most massive
cores are in a super-critical state, i.e. turbulence is not sufficient
to balance gravity, unless they are strongly magnetized
\citep{pillai2011a,kauffmann2013b}.

However, the role of magnetic fields at all relevant physical scales has
been hotly debated since over two decades
\citep{crutcher2012}. Existing magnetic field observations elsewhere
in the ISM towards lower mass local clouds or distant active star
forming regions have failed to generate a consistent picture.  This is
mainly because of systematic uncertainties like geometry, varying
spatial resolution and confused nature of of emission (protostellar vs
prestellar or combination of both). Different techniques on different
objects often yield conflicting and controversial results.

Until recently, we did not have any knowledge of the magnetic field structure or
  strength in filamentary IRDCs. This is presumably a consequence of the
relatively large time requests needed to observe emission from these
relatively faint objects (i.e., compared to regions actively forming
high--mass stars). This field has been neglected since the discovery of
these prominent structures. Therefore, magnetic field measurements is
the next important step towards understanding IRDCs and their
influence in the galactic ISM. This has recently become more urgent,
since several of the latest studies of IRDC dynamics suggests
  that magnetic fields are the dominating agent preventing the
  collapse of these clouds: internal gas motions, for example,
contain too little energy to provide significant support against
self--gravity \citep{pillai2011a,kauffmann2013b}.

There are two main techniques for studying magnetic fields in
molecular clouds. In the dust polarization technique, linear
polarization of thermal dust continuum emission and near--infrared
dust extinction is used to trace the field morphology and the
dispersion in the observed vectors to determine the field
strength. Line
splitting caused by  Zeeman effect in Zeeman sensitive molecules
like OH provides an alternate way of determining the magnetic field
strengths in molecular clouds. Both Zeeman and dust polarization signals are unfortunately very weak. Therefore, a concerted effort that would combine
both Zeeman and dust polarization techniques to understand the
magnetization in dense clouds does not exist
\citep{crutcher2012}.  Recently at the APEX telescope, the PolKa polarimeter \citep{wiesemeyer2014} has been combined with the 870$\mu$m LABOCA camera to deliver continuum polarization capabilities. Similarly, the HAWC+ camera on SOFIA has also been very recently commissioned. 
In addition, receiver and backend upgrades at the Effelsberg 100m telescope allow for sensitive Zeeman measurements. We can thus make extensive use of
all these instruments to work towards a breakthrough in understanding
magnetic fields in the dense ISM contained by IRDCs. IRDCs with minimal star formation within them
represent the unperturbed ISM where such measurements should be
ideally done. Therefore, we have launched a systematic survey (POLSTAR: POLarization in STAR forming filaments) of a sample of IRDC filaments in
both dust and line polarization with APEX, SOFIA and Effelsberg to get a unified picture of magnetic
fields  across several spatial scales (POLSTAR survey, Pillai et al. in prep.)

\section{Magnetic Fields in Filaments in the Gould Belt}
The majority of the local (within 500 pc of the Sun) filamentary molecular clouds are located in the so called Gould's Belt. Serpens-South  and Ophiucus are  two of the nearest IRDCs. Their average properties are similar to the bulk of the IRDCs in the galactic plane \citep{kauffmann2010c}.    These clouds have been extensively studied as part of the Herschel Gould Belt survey \citep{andre2010:filaments}. With the exception of the Orion molecular cloud, most filaments in the Gould Belt (GB) are low-mass star forming regions in different evolutionary stages from quiescent to active cluster forming regions. Only a small fraction (typically few percent) of the total dust emission of these filaments are expected to be polarized at mm/submm wavelengths. Therefore, deep polarization measurements were confined to small regions within such clouds.  Magnetic fields studies on large scales therefore largely relied on optical and near-IR polarimetry. Targeted studies in the B211/B213 filaments in Taurus , Serpens-South and the Musca filament show that the magnetic field  is oriented perpendicular to the dense filament and at least in some cases roughly aligned with the lower density filaments (striations) \citep{chapman2011, sugitani2011, nakamura2014, palmeirim2013, cox2016}. \citet{li2013} conducted a comprehensive study of the relation between the large scale field orientation and the filaments by analyzing the optical polarimetry catalog published by \citet{heiles2000} towards clouds covered in the Herschel Gould Belt survey. Li et al. find that filaments in the Gould Belt have a bimodal orientation  with respect to local inter-cloud medium (ICM) magnetic fields. Li et al. conclude that large scale (ICM) fields are responsible for shaping the filaments resulting in such a bi-modal distribution.

More recently, the Planck satellite has delivered a polarization map of the entire sky. While the low resolution and confusion makes the polarization data towards the inner galactic plane unusable, the small distance to the GB regions and location above the plane provides a  detailed large scale view of the magnetic field structure  of the GB clouds and the regions around them at an unprecedented level.  \citet{Planck-Collaboration2016} find that relative orientation between the magnetic field direction and the gas column density structure for 10 nearby clouds change from  parallel at low column density to perpendicular at high column densities.  Note that no evidence for a bi-modal distribution as claimed by Li et al. is discussed in the Planck study. 

All these recent observations of GB clouds  find evidence for an ordered B-field on all observed spatial scales and the observed trend in relative orientation point to a dynamically important role of B-fields. What is remarkable is that  the B-field appears to play a significant role even before the onset of star formation as evidenced by  recent optical and submm polarization study of the Polaris Flare. Polaris Flare is a translucent region devoid of any star formation activity and an optical polarization study shows that filaments are aligned along the field direction \citep{panopoulou2016}

One of the most interesting question related to filaments is its role in star formation. Star formation however happens on scales much smaller than those probed  by the magnetic field observations discussed above.  Therefore, whether the field continues to be ordered on small scales and how and where does the field undergo a change in orientation within the filaments on dense clumps and core scales is yet to be answered. In order to address this crucial question, we have observed two of the brightest filaments  in the GB region with PolKa (POLSTAR survey, Pillai et al. in prep.)

\begin{figure}
\vskip -0.cm
\hspace{-1.5cm}

\centering
$\begin{array}{ccc}
\includegraphics[angle=0,width=8.cm]{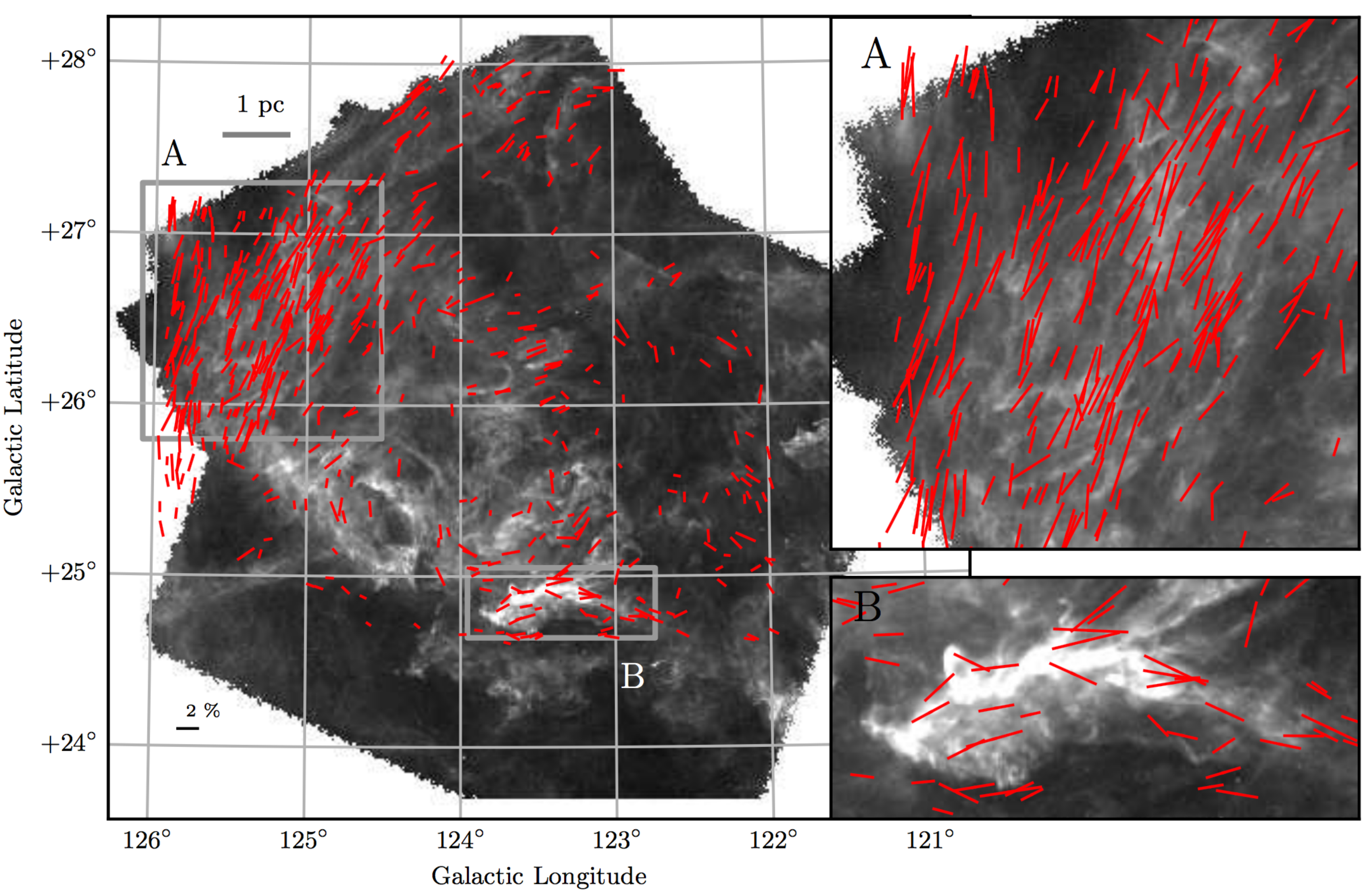} \\
\includegraphics[angle=0,width=8.cm]{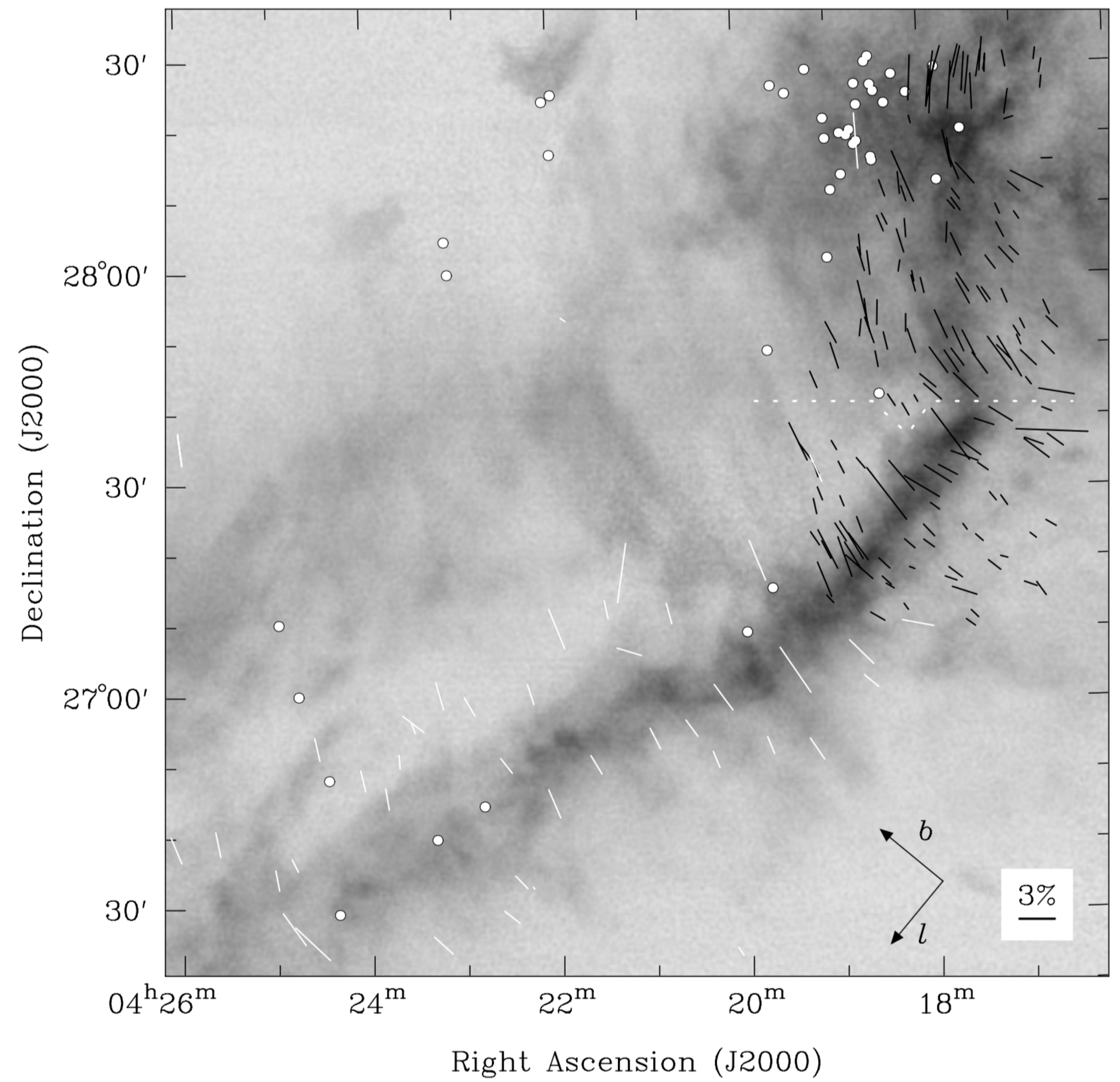}   \\
\includegraphics[angle=0,width=8.cm]{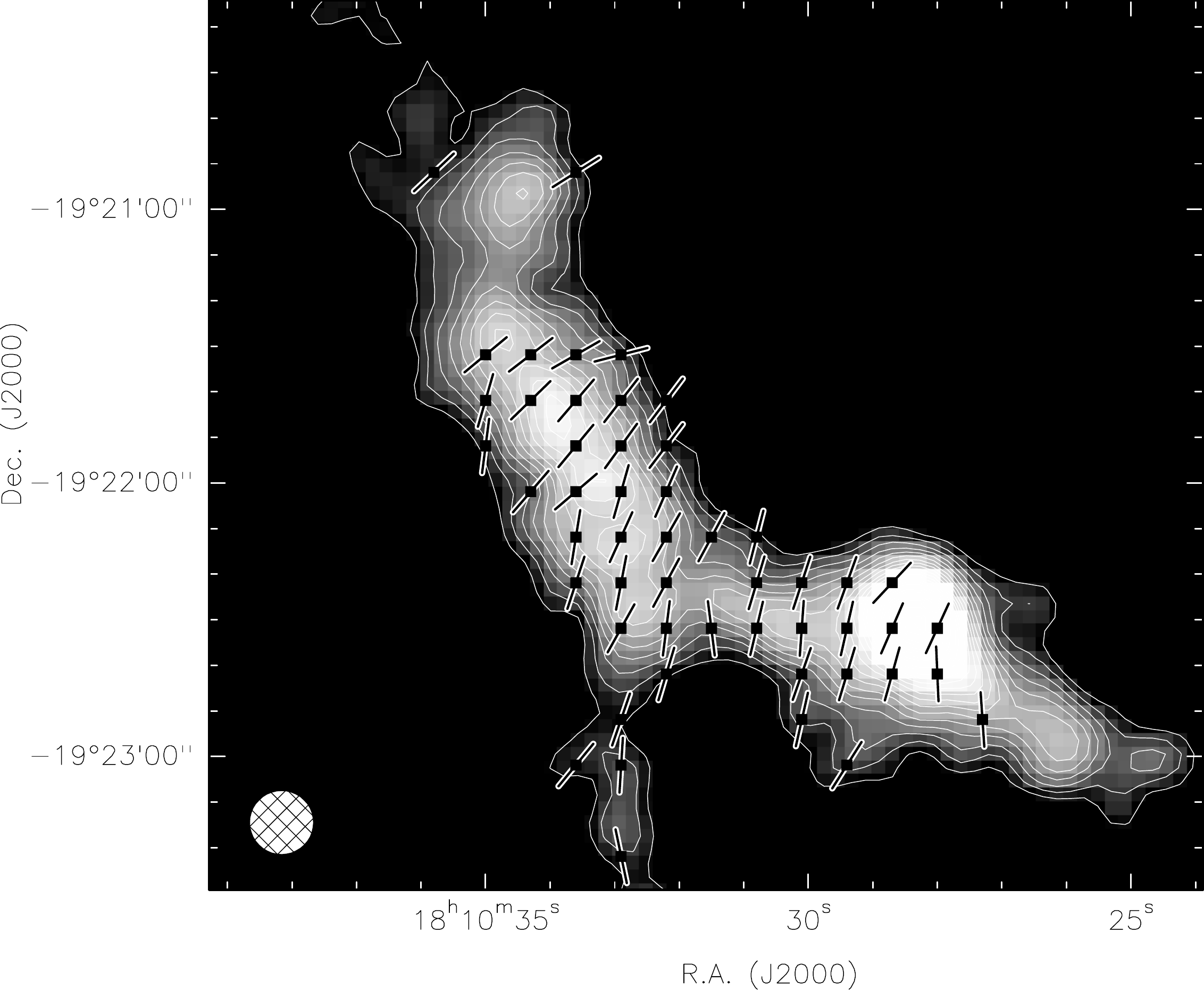} \\
\end{array}$
\caption{Magnetic field measurements of filaments in different environmental conditions and properties from diffuse to high-mass star forming. {\it Top:} Polaris Flare, a Transculent filament (greyscale image: Herschel 250$\mu$m emission). Adapted with author's permission (\citealt{panopoulou2016}). {\it Middle:} L1495/B213 filament in Taurus (greyscale image: $^{13}$CO emission ). “Optical” vectors are shown as white lines, “infrared” as black lines, and “I band” are shown as thick black lines. Embedded sources in Taurus are shown as white circles. Adapted from \citealt{chapman2011} with author's permission. {\it Bottom:} Massive star forming filament G11.11-0.12 (\citealt{pillai2015}, greyscale image and contours: SCUBA 850$\mu$m emission). Shown are magnetic field vectors obtained by rotating polarization vectors by 90\degr.}
\label{fig:4}
\vspace{-0.5cm}
\end{figure}
\section{Magnetic Fields in Filaments within the Inner Galactic Plane}
Several 1000s of filamentary IRDCs have been identified in the inner galactic plane \citep{peretto2009}. However, most of these IRDCs are not massive and dense enough to form high–mass stars and are distant analogs of GB filaments. A small number of massive and dense IRDCs may be hosts to high-mass stars \citep{kauffmann2010c}. Furthermore, a class of very long filaments in the inner galactic plane have been recently speculated to be associated with major spiral arms \citep{goodman2014, wang2016}. The densest parts of such filaments are infrared dark. Spiral arms might be responsible for triggering star formation within such IRDCs via gas accumulation and compression in spiral shocks. Spiral arm magnetic fields might also play an important role in this process \citep{shetty2006}. Do therefore the observed magnetic field properties of spiral arm massive and dense IRDCs differ from the local clouds (which should not be affected by such spiral shocks) discussed above? 

In a recent experiment using archival submm polarization data \citep{dotson2010,matthews2009}, we have reported observations of magnetic fields
in two of the most massive and pristine IRDCs in the Milky Way \citep{pillai2015}. We show
that IRDCs G11.11$-$0.12 and G0.253$+$0.016 are strongly magnetized with a
magnetic field that remarkably ordered in both clouds on sub-parsec scale, with
small angular dispersions relative to the mean cloud field. Using field
dispersion and complementary spectral line information, we also estimate the
total field strength for these IRDCs and find values of few 100 $\mu\rm{G}$. The
main dense filament in G11.11$-$0.12 is perpendicular to the magnetic field,
while the lower density filament merging onto the main filament is parallel to
the magnetic field. 

Submm dust polarization traces the  plane-of-sky projection of the actual magnetic field in the dense filament. What about the larger diffuse environment around these filamentary IRDCs?  Recently, IRDC G14.2 have been shown to be aligned perpendicular to the large scale magnetic field traced using  optical and near-infrared polarimetry techniques \citep{santos2016b}. 
Even in an active massive star forming filament like NGC6334, where feedback starts to play an important role, \citet{li2015:bfield} find that magnetic field is dynamically important.
Similar to low-mass clouds, a scenario of highly ordered field structure that is dynamically important relative to turbulence is therefore emerging. However, as can be gathered from this summary, due to the patchy nature of these studies,  no statistical significance can be placed on these findings yet. To bring this finding to a statistically firm footing, we have therefore embarked on a multi-wavelength dust polarization survey spanning near-IR (Pico dos Dias), far-IR (HAWC+, SOFIA) to submm-wavelengths (PolKa, APEX) of a large sample of nearby and distant IRDCs.  Including a large sample of filaments from quiescent to cluster-forming (low-and high-mass) will allow us to understand whether magnetic fields continue to play a dominant role in filaments as cores within such IRDCs collapse and form clusters (POLSTAR survey, Pillai et al. in prep.)

We are also conducting Zeeman experiments using Zeeman sensitive tracers in the radio (for example OH 1.6 GHz transitions) and mm (CN, \citealt{pillai2016}). Combining dust polarization and Zeeman measurements will allow us to calculate the 3D magnetic field and provide a better constraint on the total field strength. However, in spite of the increased sensitivity of instruments such as ALMA, we find that future Zeeman observations can be significantly more challenging than dust polarization measurements due to several reasons. For example, in \citet{pillai2016}, by producing a probability distribution for a large range in field geometries, we show that plane- of-sky projections are much closer to the true field strengths than line-of-sight projections. 

\section{Galactic Center Filaments}
The molecular clouds in the inner $\sim $200 pc of our galaxy  (known as the central molecular zone -- CMZ)  have significantly different properties than the Spiral Arm molecular clouds discussed above. The dense molecular cloud gas is significantly warmer, denser and more turbulent (see \citealt{Kauffmann2016c} for a recent review). Studies of filamentary molecular clouds in the CMZ of the Milky Way allow exploration of star formation under extreme conditions. In addition to being able to explore the extreme ends of the parameter space for star formation theories,  it critically informs our understanding of centers of galaxies. Thanks to multi-wavelength dedicated studies of the CMZ clouds tremendous progress has been in characterizing the large scale and sub-parsec scale properties of these clouds and the gas kinematics  \citep{longmore2013, ginsburg2016,Henshaw2016, kauffmann2016b}. This also has helped to develop dynamical models for gas distribution and star formation in the central molecular zone \citep{kruijssen2015,krumholz2016}. However, the magnetic field structure of the galactic center clouds is hardly understood. 

Previous far-IR / submillimeter polarimetry of emission from magnetically-aligned dust has revealed a large scale toroidal geometry for the field in the densest molecular component of the CMZ \citep{chuss2003}. More recent near-IR polarimetry observations provided tentative evidence for a poloidal field in the lower density environment \citep{nishiyama2010}. However, in an extreme environment such as galaxy's center, dynamical events including tidal shearing might play an important role.  Existing polarization mapping efforts are far from complete in surveying the CMZ, and suffer from limited  angular resolution in order to be able to address the relevant questions.

We have launched a systematic dust polarization survey  of the Galactic Center Molecular Clouds  with the APEX PolKa/LABOCA instrument at at 870$\mu$m. This is the first such effort to map the entire CMZ at 0.8\,parsec resolution (assuming a galactic center distance of 8.4\,kpc). This polarization survey was motivated by our pilot polarization study of the cloud G0.253+0.016 (the "Brick") using archival polarization data \citep{dotson2010}. In this study, we have shown that even in an extremely turbulent environment like the CMZ, strong magnetic fields dominate over turbulence \citep{pillai2015}.  Extending this study of magnetic fields to the rest of the CMZ clouds is  important for several reasons. First, here we can test models of molecular clouds under extreme conditions. Second, it will help us to understand the relation of the small scale cloud fields to the large scale GC field. Finally, the CMZ provides a unique arena to explore the state and evolution of molecular clouds and the role of magnetic fields in galactic centers.

\section{Summary}
Contrary to what has been the paradigm over a decade ago, it is becoming increasingly evident that magnetic fields is important in determining filamentary cloud structure. As shown in Figure 1, filaments in very different environmental conditions and properties (non-star forming and star forming) appear to show ordered fields and have a tendency to be aligned along or perpendicular to magnetic fields depending on the gas column density. Thanks to its mapping capability and high-resolution, the multi-band polarimeter HAWC+ that is currently being commissioned onboard SOFIA is one of the most powerful instrument that has the potential to provide the first large scale map of the magnetic field orientation in the dense molecular gas of filaments. The resolution allowed by HAWC+ is sufficient to disentangle the individual molecular clouds, thus informing us on how the magnetic field structures vary with galactic environment.

Needless to say, real progress in this field can only be achieved when these observations are placed in context of the suite of  numerical simulations that include gravity, turbulence and magnetic fields. A full 3D analysis of the filamentary structures formed in MHD  simulations with different initial conditions, their gas kinematics and magnetic field structure (derived using dust radiative transfer codes that include realistic dust grain alignment mechanisms \citealt{reissl2016})  at different time steps should be the next step.

\section*{Acknowledgments}
\vspace{-0.3cm}
T.P. acknowledges support from the \emph{Deut\-sche For\-schungs\-ge\-mein\-schaft, DFG\/} via the SPP (priority program) 1573 ‘Physics of the ISM’.

\end{document}